\documentclass{mem}
\usepackage{natbib}\usepackage{txfonts}\usepackage{balance}
\usepackage{graphicx}
\usepackage[a4paper]{hyperref}
\idline{75}{282}
\begin{document}
\def\teff{$T\rm_{eff }$}
\def\kms{$\mathrm {km s}^{-1}$}

\title{
Testing 3D solar models against observations
}

   \subtitle{Centre-to-limb variations of oxygen lines, spatially-resolved line formation and probing for departures from LTE}

\author{
T. M. D. \,Pereira\inst{1,2} 
\and M. \, Asplund\inst{3}
\and D. Kiselman\inst{2}
          }

  \offprints{T. M. D. Pereira, \email{tiago@mso.anu.edu.au}}

\institute{
Research School of Astronomy and Astrophysics, Australian National University, 
Cotter Rd., Weston, ACT 2611, 
Australia,
\and
The Institute for Solar Physics of the Royal Swedish Academy of Sciences, 
AlbaNova University Center, 106 91 Stockholm, 
Sweden
\and
Max-Planck-Institut f\"ur Astrophysik, 
Postfach 1317, D--85741 Garching b. M\"unchen,
Germany
}

\authorrunning{Pereira et al.}

\titlerunning{Testing 3D solar models against observations}

\abstract{
We present results from a series of observational tests to 3D and 1D solar models. In particular, emphasis is given to the line formation of atomic oxygen lines, used to derive the much debated solar oxygen photospheric abundance. Using high-quality observations obtained with the Swedish Solar Telescope (SST) we study the centre-to-limb variation of the O\,\textsc{i} lines, testing the models and line formation (LTE and non-LTE). For the O\,\textsc{i} 777 nm triplet, the centre-to-limb variation sets strong constraints in the non-LTE line formation, and is used to derive an empirical correction factor ($S_{\mathrm{H}}$) to the classical Drawin recipe for neutral hydrogen collisions. Taking advantage of the spatially-resolved character of the SST data, an additional framework for testing the 3D model and line formation is also studied. From the tests we confirm that the employed 3D model is realistic and its predictions agree very well with the observations.
}
\maketitle{}

\section{Introduction}

Three-dimensional, time dependent hydrodynamical simulations of stellar atmospheres represent a paradigm change in the modeling of stellar atmospheres \citep{Asplund2005}. Unlike their one-dimensional counterparts, they treat convection self-consistently without the need for broadening parameters such as micro and macroturbulence. Their realistic description of the motions and velocities in the photosphere leads to an excellent agreement between the predicted and observed line shapes and shifts \citep{Asplund2000}.

\citet{AGS05} have revised the solar chemical composition, using photospheric lines the 3D solar model of \citet{SteinNordlund1998}. This revision implied a lower metal content, in particular C, N and O, which caused significant in the agreement of solar interior models with results from helioseismology \citep{Bahcall2005,Basu2008}. The new chemical composition of \citet{Asplund2009} results in a slightly higher solar metalicity, but \emph{per se} is not enough to reconcile the solar interior models with helioseismology. To trust the results from the 3D models, more testing is needed. Our objective is to test the 3D model used by \citet{Asplund2009}, in particular for the oxygen lines, determinant in the downward revision of the solar abundances. This contribution summarizes the findings of \citet{Pereira2009b} and \citet{Pereira2009a}, where the centre-to-limb variation and spatially-resolved line formation of oxygen lines is tested.

\section{Observations and photospheric models}

Using the Swedish 1-m Solar Telescope (SST) with the TRIPPEL spectrograph, we obtained high spatial and spectral resolution observations at several positions in the solar disk. The lines observed include five atomic oxygen lines at 615.81~nm, 630.03~nm and the three lines around 777~nm. Stray light in the telescope was estimated and corrected for. The observations are detailed in \citet{Pereira2009b}. 

We test these observations against a new 3D model atmosphere (Trampedach et al. 2009, in preparation). It was computed with the \textsc{stagger} MHD code \citep{NordlundGalsgaard1995,GudiksenNordlund2005}, and consists of a $240^3$ grid with a physical size of 6$\times$6$\times$4\,Mm. It includes an improved treatment of radiation, with a 12-bin multi-group opacity binning scheme. For the line formation calculations the original simulation was interpolated to a 50$\times$50$\times$82 grid to save computing time. The simulation snapshots used here cover $\approx$45 minutes of solar time.

In addition, we also compare the centre-to-limb variation predicted by two 1D models: the semi-empirical model of \citet{HM1974} and the LTE, line-blanketed \textsc{marcs} model \citep{MARCS2008}.

Oxygen line formation calculations were done in LTE and NLTE, using our LTE code and the \textsc{multi3d} code \citep{Botnen1997,Botnen1999,Asplund2003a}. A 23-level model atom was used \citep[see details in][]{Pereira2009a}. For the O\,\textsc{i} 615.81~nm and [O\,\textsc{i}] 630.03~nm lines we show the LTE results, because the NLTE effects are weak in those lines. For the O\,\textsc{i} 777~nm lines NLTE calculations were carried out for several values of $S_\mathrm{H}$, a scaling factor for the collisions with H\,\textsc{i} from the classical formul\ae{} \citep{Drawin1968,Steenbock1984}. Nearby blends were included in the calculations of the 615.81~nm line (in particular, CN and C$_2$ molecular blends) and the 630.03~nm line (in particular, the Ni\,\textsc{i} blend).

\begin{figure} 
  \centering
  \includegraphics[width=0.5\textwidth]{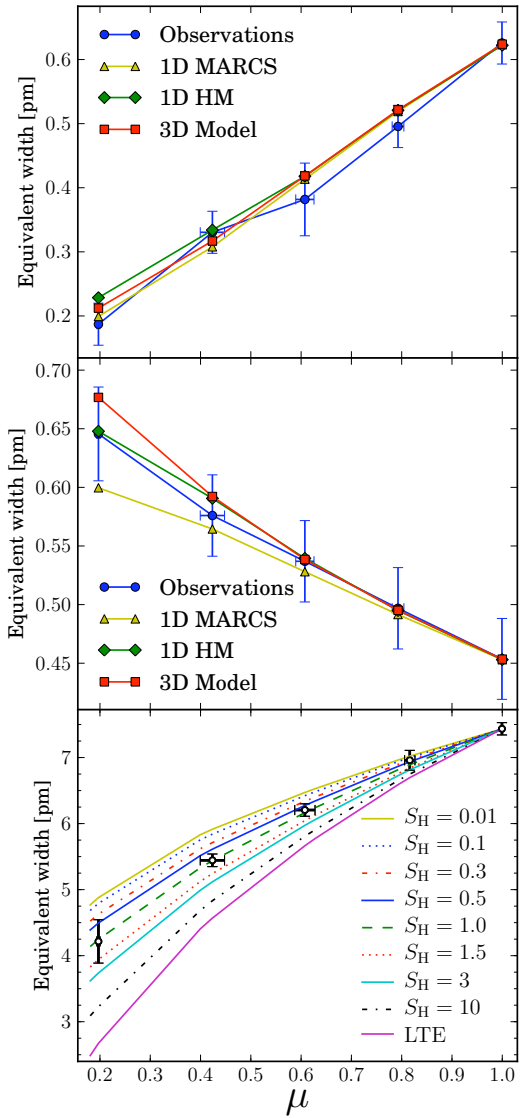}
  \caption{Centre-to-limb variation of the oxygen lines. \emph{Top:} O\,\textsc{i} 615.81~nm line (including blends). \emph{Middle:} [O\,\textsc{i}] 630.03~nm line (including blends). \emph{Bottom:} O\,\textsc{i} 777.41~nm line, only for the 3D model and for different scalings of the H\,\textsc{i} collisions.}
  \label{fig:clv_eqw}
\end{figure}

\section{Centre-to-limb variation}

In Fig.~\ref{fig:clv_eqw} we show the centre-to-limb variation of some of the lines. For the 615.81~nm and 630.03~nm lines the 3D model predictions agree well with the observations. For these lines there is only a small difference between the different models.

For the O\,\textsc{i} 777~nm lines, the efficiency of the collisions with neutral hydrogen in the NLTE calculations is unknown. Following \citet{CAP2004}, we use the centre-to-limb variation of these lines to derive an empirical estimate of $S_\mathrm{H}$, the multiplier factor by the classical collision rates. We show only one representative line, and only for the 3D model. Of the $S_\mathrm{H}$ values used, we find that $S_\mathrm{H}=1$ gives the best agreement with the observations. In \citet{Pereira2009a} a further refinement of this value is made, and the best fitting  value is found to be $S_\mathrm{H}\approx 0.85$. For these lines, one can very clearly rule out LTE as an acceptable approximation. The 777~nm results for the 1D Holweger--M\"uller model are very similar to those of the 3D model. The \textsc{marcs} model fails to reproduce the observations, regardless of the $S_\mathrm{H}$ used. 

\section{Spatially-resolved line formation}

\begin{figure*} 
  \centering
  \includegraphics[width=\textwidth]{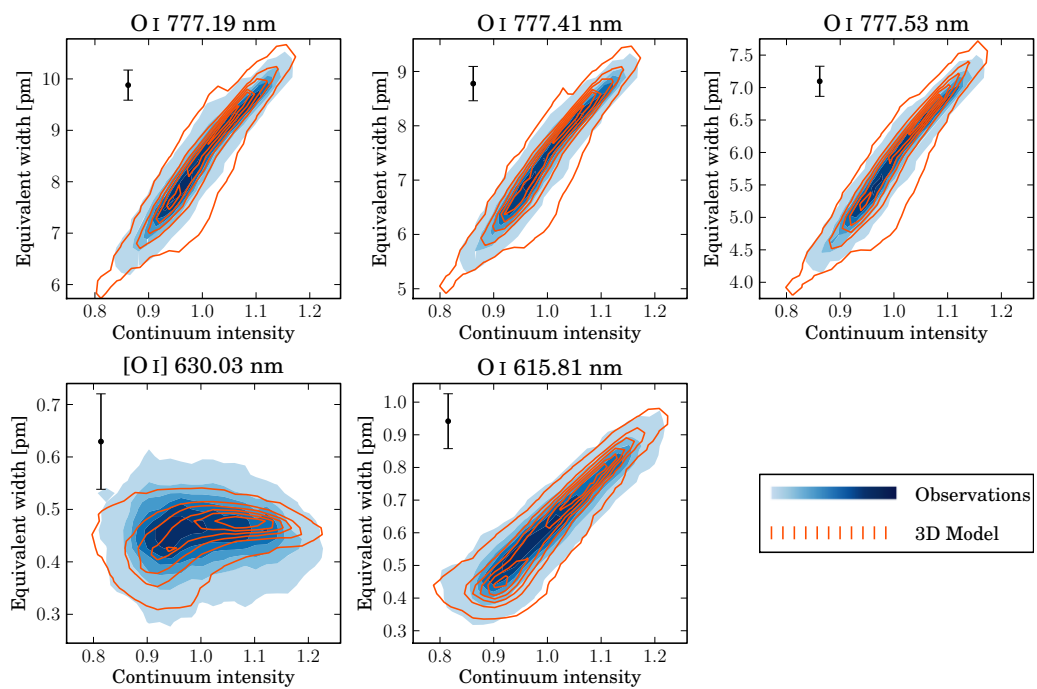}
  \caption{Distribution of equivalent widths over the solar granulation at disk-centre, for five oxygen lines. The 3D model results have been degraded to account for the atmospheric turbulence and the telescope's resolution. For each spatially-resolved spectrum the equivalent width has been computed and the resulting histogram of points is represented by the contours, as a function of the normalized local continuum intensity. The line profiles of the O\,\textsc{i} 777 nm lines were computed in NLTE, for $S_\mathrm{H}=1$. For the other two lines LTE was assumed.}
  \label{fig:spatial_resolved}
\end{figure*}

In Fig.~\ref{fig:spatial_resolved} we show the distribution of the equivalent widths in the disk-centre granulation. The results from the 3D model have been convolved with a point-spread function to account for the atmospheric smearing and the telescope's resolution, and with a Gaussian to account for the spectrograph's instrumental profile. The oxygen abundance was adjusted so that the predicted mean temporal and spatial equivalent widths matched the observed. 

For the O\,\textsc{i} 777~nm lines the NLTE results are shown, with $S_\mathrm{H}=1$. It is comforting to find that $S_\mathrm{H}=1$ gives the best agreement also here.

\section{Conclusion}

We looked at the centre-to-limb variation and spatially-resolved line formation of oxygen lines, comparing new observations with the 3D model. These lines are an important test of the model, because oxygen is very relevant in the recent revisions of the solar chemical composition.

Overall there is a very good agreement between the predictions from the 3D model and the observations. Both in the centre-to-limb variation and the spatially-resolved line formation for oxygen. These results give us confidence that the 3D model is realistic and appropriate to derive the solar chemical composition, in particular the oxygen abundance.

\begin{acknowledgements} 
TMDP acknowledges financial support from Funda\c c\~ao para a Ci\^encia e Tecnologia (reference number SFRH/BD/21888/2005) and from the USO-SP International Graduate School for Solar Physics under a Marie Curie Early Stage Training Fellowship (project MEST-CT-2005-020395) from the European Commission. This research has been partly funded by a grant from the Australian Research Council (DP0558836). The Swedish 1-m Solar Telescope is operated on the island of La Palma by the Institute for Solar Physics of the Royal Swedish Academy of Sciences in the Spanish Observatorio del Roque de los Muchachos of the Instituto de Astrofísica de Canarias. 
\end{acknowledgements}


\end{document}